\documentclass[prc,aps,preprint]{revtex4}

\usepackage{graphicx}

\begin{document}

\title{How Relativistic Heavy Ion Collisions\\
       Can Help Us Understand the Universe}
\author{Berndt M\"uller}
\affiliation{Department of Physics, Duke University,
             Durham, NC 27708-0305, USA}
\author{Dinesh K. Srivastava}
\affiliation{Physics  Group, Variable Energy Cyclotron Center,
              Kolkata 700064, India}

\date{\today}
\begin{abstract}
We discuss the anthropic principle and its implications for our existence
and the physical laws which govern the universe. Several amazing coincidences 
which provide conditions necessary for creation of life suggest that the 
``laws of nature'' are not uniquely determined. The idea that our universe 
is only one among a multitude of universes with different physical laws, 
as predicted by the theory of chaotic cosmic inflation, provides a logically
simple, but speculative resolution of the anthropic principle. An important
insight of modern quantum field theory is that the physical laws are not
only determined by symmetry principles, but also by the nature of the vacuum
state. Experiments involving collisions of relativistic heavy ions provide 
the clearest tests of the hypothesis that properties of particles and forces 
can be modified by a change in the vacuum state. We outline the goals of
these experiments and briefly review their current status.
\end{abstract}
\maketitle

\section{Introduction}

Understanding the large-scale structure of the universe and the
fundamental laws governing natural phenomena have been dual goals
of science since its origins in ancient Greece. Over the past
century, remarkable progress in physics and astronomy has brought
us increasingly closer to the goal, in the words of Goethe's 
{\em Dr.~Faustus:} ``... zu erkennen, was die Welt im Innersten 
zusammenh\"alt.''\footnote{``... to discern how the world is 
put together at its core.''}  Big Bang cosmology and the Standard
Model of particle physics combine into a nearly flawless tapestry 
describing the evolution of our universe from the age of a few 
seconds to our present time. As awe inspiring as this picture is, 
it contains a few blurred patches and some unexplained, but 
highly intriguing features:
\begin{itemize}
\item
We do not know the nature of 97\% of the mass and energy content
of the universe;
\item
We do not understand why the cosmological constant appears to
be nonzero, yet many orders of magnitude smaller than dimensional
arguments would suggest;
\item
We do not understand which principle determines the values of
the more than 20 ``fundamental'' parameters in the Standard Model,
the particle masses, mixing angles, and coupling constants;
\item
Yet, the smallness of the cosmological constant and subtle, 
mysterious relationships among the fundamental constants, are
essential determinants of the possibility of the emergence of
life in the universe, and hence of our own existence.
\end{itemize}
The curious fact that our existence is dependent on several
remarkable coincidences among the physical constants and the
detailed properties of atoms and nuclei is known as the {\em 
anthropic cosmological principle}, or anthropic principle in short
\cite{Ca74,CR79,Dy79,Da82,BT86}. It has a peculiar place in modern 
science, embraced by some leading scientists of our day and
abhorred by others \cite{CaseConf}. Most scientists believe that 
the fundamental laws of nature have an objective origin and are 
not human constructs or mere accidents. On the other hand, the
numerical coincidences in the natural laws appear so contrived 
that they are hard to reconcile with our predilection against
fine tuning. 

Yet, this reconciliation is not impossible. Science has progressed 
a long way from the creation myths of ancient cultures, which
explained the fecund environment on our planet as the creation
of a supernatural being. We now understand that life on Earth 
has been formed by evolutionary forces under boundary conditions, 
which appear to us as similarly fortuitous: The sun is a star of just 
the right mass to provide life sustaining energy for billions of
years, long enough to support the emergence of intelligent life. 
The planet Earth has the right size and just the right distance 
from the sun to retain large amounts of liquid surface water. 
The initial intensity of asteroidal impacts onto the Earth was 
large enough to leave our planet with a thin crust, facilitating 
continuing rearrangement of the continents, and to set it into 
rapid rotation, leading to a moderate temperature gradient between 
day and night -- all due to the giant impact that created the moon 
some 4.5 billion years ago \cite{Canup}. The impact activity then 
waned sufficiently, but not completely, to permit the evolution 
of higher forms of life by producing rare moments of large upheaval 
which eliminated less efficient competitors, e.g.\ the dinosaurs, 
without destroying life in its entirety.

We now know that, no matter how special these circumstances are,
the universe is large enough and old enough for them to arise 
with reasonable likelihood on 
one of the planets in some solar system in some galaxy. According
to rough estimates, the visible universe contains some $10^{23}$
stars, which means that if the combined probability for all the 
coincidences and fortuitous circumstances required for the emergence 
of intelligent and civilized life is not less than $10^{-23}$, 
then it will appear somewhere in the cosmos. This consideration
also allows us to address the question whether we are ``alone''
in the universe. Of the $10^{11}$ or so solar systems in our
galaxy, how many harbour intelligent life? This question is 
answered by Drake's equation \cite{DrakeEq}
\begin{equation}
N_{C} = R^* \times p_E \times p_L \times p_I \times L ,
\end{equation}
where $N_C$ is the number of contemporaneous civilizations in
our galaxy, $R^*$ is the rate of formation or sun-like stars,
$p_E$ is the probability for such a star to have an Earth-like
planet, $p_L$ the probability for life to form on such a planet,
$p_I$ the probability for intelligent life to emerge, and $L$ 
the average lifetime of a civilization. We do not know the
various components of this relation with precision, but inserting
reasonable estimates and multiplying the result by the number
of galaxies in the visible universe ($10^{12}$), we conclude 
that the existence of another high civilization somewhere else 
in the universe is quite likely.

We could actually turn the argument on its head and reason that
the sheer improbability of the fortunate coincidences permitting
our presence on Earth demands the existence of a very large
number of planets with varying physical parameters and different 
history. This would convert the apparent improbability of our
existence into likelihood and, of course, our home planet would
by necessity be one that can sustain intelligent life. Given the
vast number of planets required to make this reasoning viable, 
it would be an unlikely coincidence if that number were just
sufficient to produce a single case of intelligent life. Hence,
by means of probability arguments, we would surmise that the
universe contains many planets inhabited by intelligent beings,
though not necessarily all at the same time, if civilizations
do not last for periods of cosmological duration. 

\begin{figure}[tb]   
\begin{center}
\includegraphics[width=0.7\linewidth]{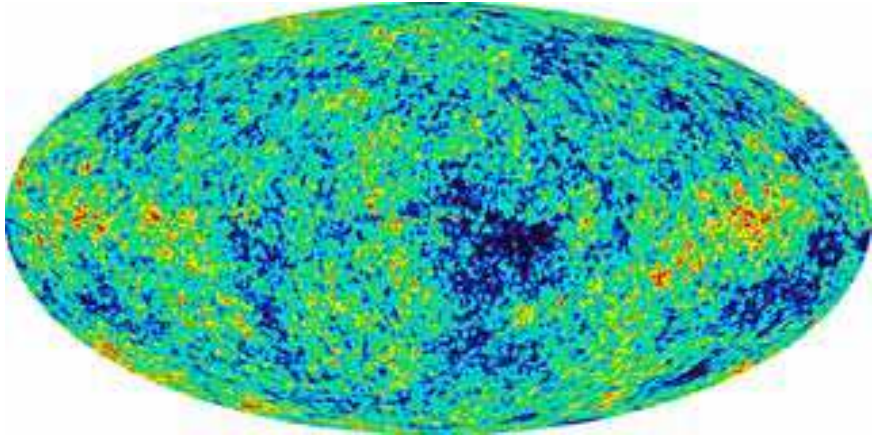}
\end{center}
\caption{Temperature fluctuations in the cosmic background radiation
         at the $10^{-5}$ level (from the WMAP experiment).}
\label{fig1}
\end{figure}

But, if this probability argument is correct, could we expect 
ours to be a ``typical'' civilization? This speculation would 
only be justified, if we had some idea of the parameter space 
defining inhabitable planets including their likelihood of 
formation, {\em and} would know that the Earth falls into the 
more probable regions of this space. Absent this knowledge, or 
some statistical information about other civilizations, there 
is no {\em a priori} justification of the expectation that we 
are typical or average on a cosmic scale. In view of the rapid 
progress in the search for planets orbiting other stars, we may
soon be able to give a first tentative answer to this question,
but we are not quite able to do so yet.

\section{Cosmic Coincidences}

We know that our universe was formed about 14 billion years
ago in a state of very high temperature and has since expanded
and cooled to a temperature of 2.73 K. A careful analysis of the
abundance of light elements tells us that the universe has grown 
in size by at least 10 orders of magnitude from its early state, 
and probably by many more. The magnificently detailed map of the
cosmic background radiation assembled by the WMAP satellite 
(shown in Fig.~\ref{fig1}) and
other experiments shows that the universe is isotropic and 
homogeneous on a large scale, with relative thermal fluctuations at 
the $10^{-5}$ level. The data also show that the large-scale geometry
of the universe is flat, or nearly flat, and that all known forms
of matter constitute no more than 3\% of its mass-energy content.

Einstein's equations of general relativity tell us that the
present approximate flatness implies that the universe was flat
with a precision of better than $10^{-20}$ during the period 
of cosmic nucleosynthesis. Following our inbred abhorrence of 
fine-tuning requirements, we are led to postulate that the 
universe is, on cosmic scales, absolutely flat. One consequence
of this assumption is, in a somewhat loosely defined sense, that 
the positive energy of all matter in the universe, including the 
kinetic energy involved in the expansion, is precisely balanced 
by the negative energy of the attractive gravitational potential. 
In other words, the total energy content of the universe is zero, 
or at least very small compared with its various components taken 
separately. This suggests that the universe may have been created 
by an event of microscopic dimensions, maybe some kind of quantum 
fluctuation.

\begin{figure}[tb]   
\begin{center}
\includegraphics[width=0.7\linewidth]{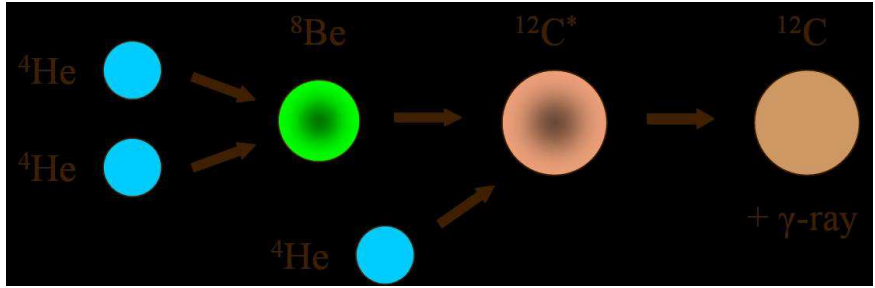}
\end{center}
\caption{Fusion chain involving three $^4$He nuclei leading to
         the formation of a $^{12}$C nucleus. Due to the instability
         of the $^8$B nucleus, the reaction rate is very sensitive
         to the location of the compound nucleus resonance in $^{12}$C.}
\label{fig2}
\end{figure}

Let us review some of the remarkable coincidences that have
made life in our cosmos possible. Maybe the best known example
is the presence of a narrow resonance state in the excitation 
spectrum of $^{12}$C at 7.65 MeV. The location and width of this 
state control the fusion rate of three $^4$He nuclei into a 
$^{12}$C nucleus (see Fig.~\ref{fig2}), and thus are determining 
factors of the abundance 
of carbon and heavier elements in the universe. The existence of 
this resonance is so essential, that it was predicted by Hoyle 
\cite{Ho54} in order to explain the cosmic carbon abundance. 
The energy of this state depends sensitively on the strength of 
the nucleon-nucleon interaction, as well as on the proton and 
neutron masses. Microscopic calculations \cite {OPC99} have shown 
that abundant production of $^{12}$C in stellar fusion requires 
a fine-tuning of the strength of the nuclear force to the 
precision of $\pm 4$\%.

Another salient example of apparent fine-tuning is the neutron --
proton mass difference $\Delta m$, which is a subtle balance of
the difference between the masses of the $u$- and $d$-quarks and
the electromagnetic self-energies of the proton and neutron. 
The mass difference $\Delta m/m \approx 1.8\times 10^{-3}$ controls 
both, the lifetime of the free neutron and the regions of stability
against beta-decay of atomic nuclei. If the neutron were only
0.15\% lighter or 0.3\% heavier, the stablity and cosmic abundance
of nuclei important for life would be seriously affected.

There is a delicate balance between the rate of energy production 
in the interior of stars (governed by the density and temperature 
of the stellar core), the rate of energy transport to the surface 
(determined by the stellar radius and the temperature profile), 
and rate of energy radiation from the surface (governed by the
surface temperature). The existence of stable, very long-lived
stars with significant energy output, like our sun, relies on the
near coincidence of two very large dimensionless quantities
\cite{Ca74}:
\begin{equation}
G m_p^2 / \hbar c \approx 6\times 10^{-39}
  \approx \alpha^{12} (m_e/m_p)^4 ,
\label{eqstar}
\end{equation}
where $G$ is Newton's gravitational constant, $\alpha\approx 1/137$
the electromagnetic fine-structure constant, and $m_e, m_p$ denote
the electron and proton mass, respectively. Since the Standard Model 
of particle physics gives us no clue about the relation of any of 
these four constants to each other, this coincidence is truly amazing. 
It is very difficult to imagine how a fundamental theory, which would 
predict these relations, could lead to (\ref{eqstar}) except by sheer 
coincidence.

Finally, a recently much discussed example of a comic coincidence 
is the observed value of the cosmological constant \cite{CCrefs}
\begin{equation}
\Lambda \approx (2\times 10^{-3} {\rm eV})^4 ,
\label{CC}
\end{equation}
which is more than 120 orders of magnitude smaller than the
``natural'' scale set by the Planck mass $M_{\rm P}$ cut-off of 
quantum field theories. If $\Lambda$ were to exceed the observed 
value by a factor 200 or more, the universe would not have 
gone through the slow period of expansion that is required for 
the formation of large galaxies, and it is unlikely that life could 
have developed \cite{We87}. Even in supersymmetric models of grand 
unification the cosmological constant is at least 60 orders of 
magnitude larger than observed \cite{SUSY}. 

The anthropic principle has been invoked to explain why the laws 
of nature and the numerical constants in these laws have the 
specific forms or values that are observed in nature. But this 
``explanation'' comes at a cost: it postulates that, in some 
unspecified way, the universe at large ``knows'' about our existence. 
Who ordered those values? Do we need a purposeful creator of the cosmos, 
after all? 

Scientific reasoning provides for a way out of this dilemma. 
Our argument that the planet Earth is special, but its existence 
is nonetheless probable because of the vast multitude of planets, 
leads us to the hypothesis that our universe may be a special, 
but probable representative of a multitude of universes, each 
with different properties. This hypothesis has several immediate
implications, which may initially appear rather radical:
\begin{itemize}
\item
Our universe is only one among many others;
\item
The constants of nature and cosmological parameters must have 
different values in different universes;
\item
We happen to live in a universe where the laws of nature are 
conducive to the formation of intelligent life.
\end{itemize}
A corollary of the second statement is that not all constants
of nature can be derived from some fundamental theory (\ref{eqstar}). 
Examples are the cosmological constant, the ratio of the $u$- and 
$d$-quark masses \cite{Ho01}, and some constants entering into the 
relation.\footnote{It is noteworthy that this argument also provides 
a means to invalidate this strong form of the anthropic principle, 
e.~g.\ by showing that the cosmic coincidences are, in fact, 
predicted by some theory underlying the Standard Model.}

\section{The Role of the Vacuum}

The universality and immutability of the fundamental laws of 
nature in our universe is well established by observations. 
How can this fact be reconciled with the concept of many 
universes governed by quantitatively different laws? A possible
resolution of this paradox relies on the special role of the
vacuum state in modern quantum field theory. Quantum mechanics
dictates that even ``empty'' space is not empty, but rather
filled with quantum fluctuations of all possible kinds. The
uncertainty principle, $\Delta E \cdot \Delta t \geq \hbar$, 
is inconsistent with the notion of an absolutely empty region 
of space-time. The physical reality of the quantum fluctuations
can be measured by changing the geometry of a volume, leading
to the so-called Casimir force \cite{Cas48}: two conducting 
metal plates separated by a distance $a$ attract each other 
with a force per unit area proportional to $a^{-4}$, which 
originates from the geometry dependence of the fluctuating 
modes of the electromagnetic field between the plates.

\begin{figure}[tb]   
\begin{center}
\includegraphics[width=0.25\linewidth]{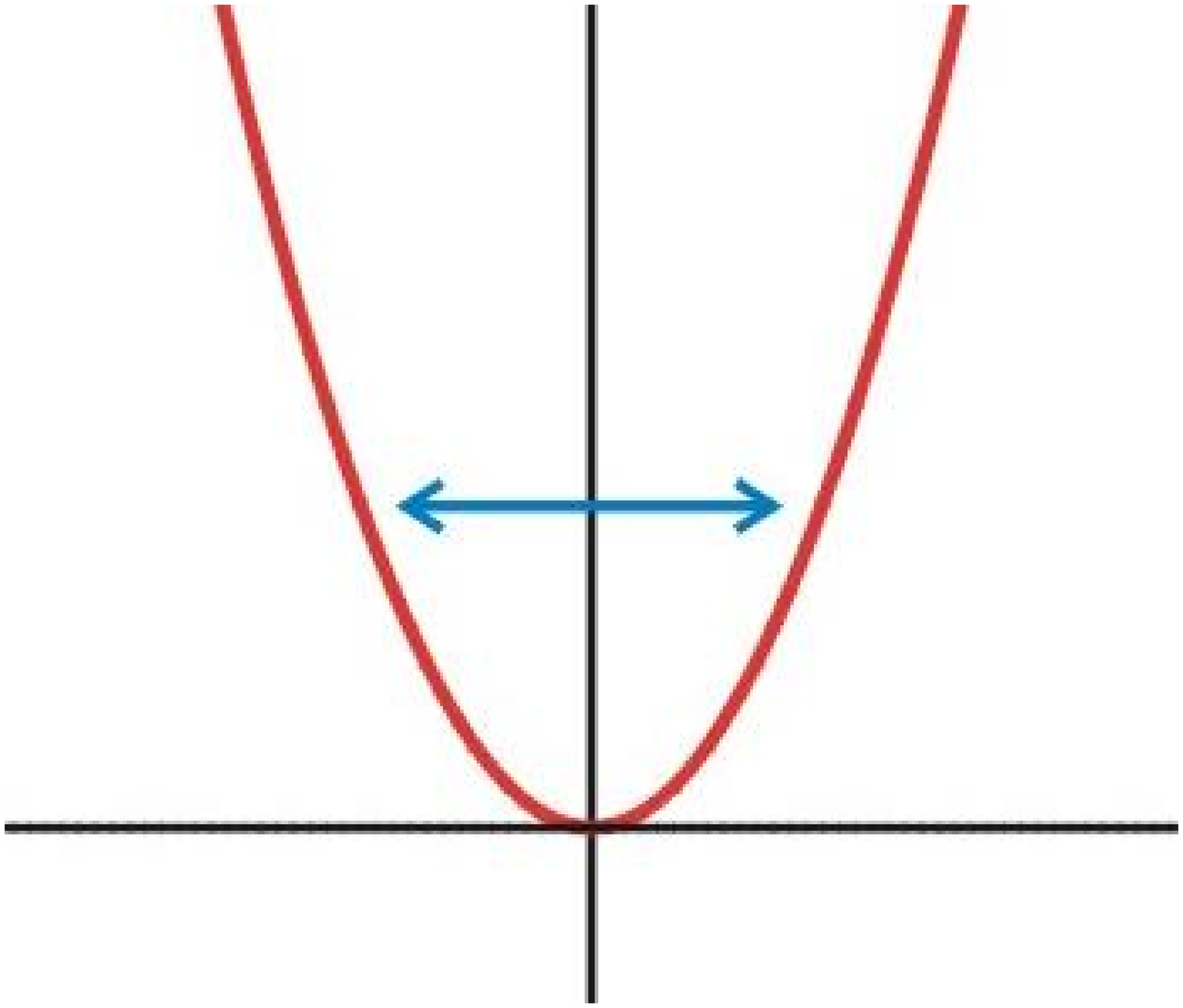}
\hspace{0.05\linewidth}
\includegraphics[width=0.26\linewidth]{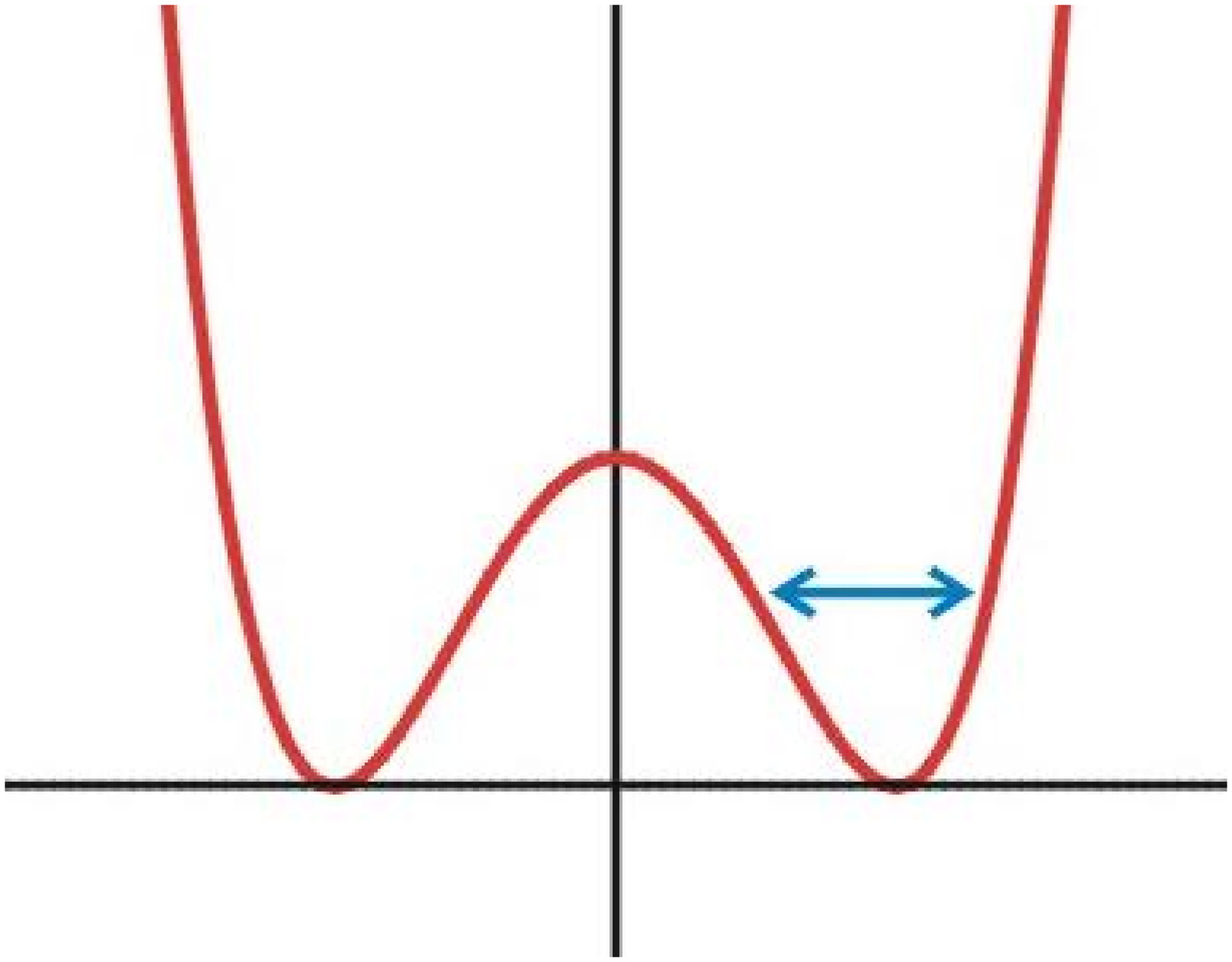}
\hspace{0.05\linewidth}
\includegraphics[width=0.25\linewidth]{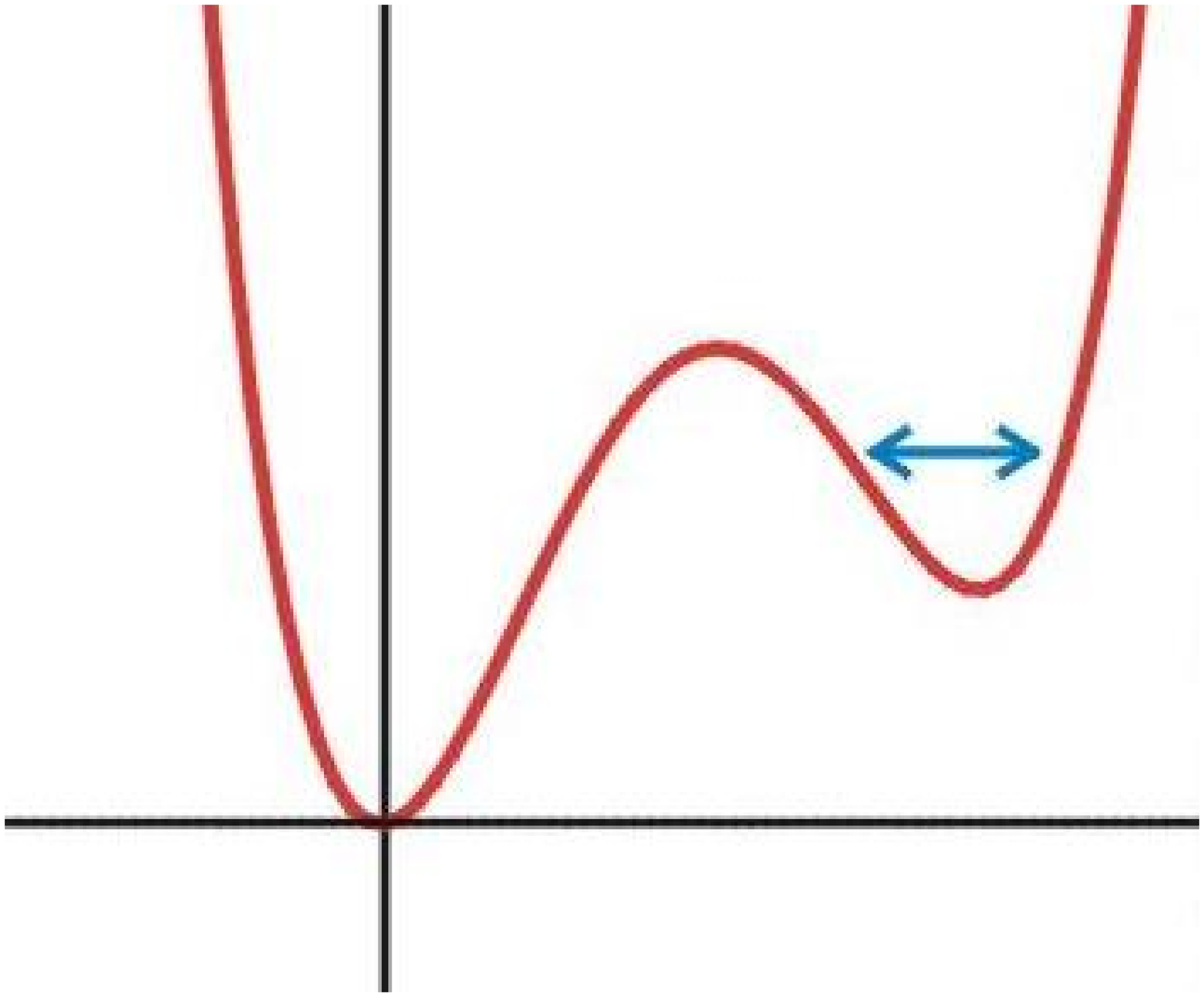}
\end{center}
\caption{Possible shapes of the potential energy as function of 
         the field strength. A minimum corresponds to a vacuum
         configuration. From left to right: trivial vacuum, Higgs 
         vacuum, and ``false'' vacuum.}
\label{fig3}
\end{figure}

Quantum field theory distinguishes three different types of
vacua: The trivial vacuum, characterized by fluctuations of
the field around zero (see Fig.~\ref{fig3}a), the Higgs 
vacuum, characterized by fluctuations around a nonvanishing
value of the field called the vacuum expectation value (see
Fig.~\ref{fig3}b), and the false vacuum, where the field
fluctuates around a metastable minimum of the potential 
energy curve (see Fig.~\ref{fig3}c). Because the false vacuum
is only metastable, it eventually decays, releasing the energy
difference between the local minimum and the true minimum of
the field potential as thermal energy.

The Standard Model uses the Higgs vacuum to generate masses
for the quarks and leptons, as well as for the gauge bosons
of the weak interaction. The vacuum expectation value of the
Higgs field, $\langle\phi\rangle = 246$ GeV, is uniquely 
determined in the Standard Model; the quark and lepton masses
differ from one another due to the different strength of the
coupling of each fermion field to the Higgs field. In models
of grand unification, the coupling strengths of the electroweak
and strong interactions themselves depend on the expectation
value of another Higgs field, which breaks the underlying
fundamental symmetry. In addition, the QCD vacuum contains
quark and gluon condensates
\begin{equation}
\langle{\bar q}q\rangle = (235 {\rm MeV})^3 , \qquad
\langle g^2G_{\mu\nu}G^{\mu\nu}\rangle = (840 {\rm MeV})^4 ,
\end{equation}
which generate additional dynamical masses for the three 
light quark flavours $u, d, s$.

Modern cosmology makes use of the false vacuum to explain the 
large size and homogeneity of our universe, as well as the 
fact that it is filled with thermal radiation. Because gravity 
couples to the full energy-momentum tensor $T^{\mu\nu}$, and 
the tensor for the false vacuum has a Lorentz invariant 
structure with negative trace $T^\mu_\mu = v g^\mu_\mu = - 2v$, 
the false vacuum effectively acts as a source of antigravity. 
In the context of the equations of general relativity, it leads 
to a law of exponential inflation of the radius of the universe:
\begin{equation}
dR/dt = R \sqrt{8\pi G v/3} \equiv H R  .
\end{equation}
Inflationary cosmology \cite{Gu81,Li82,AS82} solves, at once, 
several of the fine-tuning problems of cosmology by stretching 
away any initial deviations from flatness, isotropy, and 
homogeneity. It also explains why the universe was born hot: 
The thermal background radiation is simply the thermalized 
remnant of the latent heat contained in the false vacuum that 
drove the cosmic inflation. In order to explain the observations, 
the linear size of the universe must have grown by at least 26 
orders of magnitude, and possibly much more, during the early 
inflationary period \cite{Gu00}.

Remarkably, the inflationary cosmological model contains in 
itself an additional unexpected, philosophically important 
insight. In most models of inflation, an infinite sequence 
of universes are created once inflation has started 
\cite{St83,Vi83}. The reason for this astonishing property is 
that the volume of space filled by the excited vacuum state 
inflates faster than the excited state can decay. In other 
words, as the false vacuum decays in one region of space, 
creating a new universe, the remaining space still filled with 
false vacuum continues to grow exponentially, allowing for the
formation of other bubbles in which the excited vacuum decays, 
each one causally disconnected from all others. This process of 
``eternal inflation'' results in a scenario where an unlimited 
number of universes develop in isolation, separated by rapidly
stretching regions of space still filled with the false vacuum.
Numerical simulations of this scenario yield, at any given moment, 
the picture of a fractal ``multiverse'' filled with isolated
bubbles that have set out on their Big-Bang evolution at various 
times in the past \cite{VVW00}.

The multiverse hypothesis can form as the basis of a solution to 
the problem of cosmic coincidences, if the true vacuum state is
strongly degenerate, allowing the Higgs fields to take on
many different values, either at random or in response to
subtle initial or boundary conditions. The quantitative laws
of physics would then be different in each bubble universe,
because they contain different vacua. The anthropic principle 
then simply implies that we live in one of those universes that 
are conducive to the formation of intelligent life.

\section{Probing the Vacuum}

How can one probe the validity of these highly speculative 
concepts derived from the fusion of quantum field theory and
cosmology? Scientists are proceeding along different avenues
toward this goal:
\begin{enumerate}
\item
Astronomers measure the vacuum fraction of the energy balance 
of our universe. This ``dark energy'' has been determined to
make up about 70\% of the cosmic energy content.
\item
High energy physicists search for the Higgs boson in particle 
collisions at the Fermilab Tevatron and soon at the CERN Large
Hadron Collider (LHC).
\item
Nuclear physicists create conditions, similar to those after 
the Big Bang, in collisions of nuclei at the Relativistic Heavy
Ion Collider (RHIC) at Brookhaven National Laboratory. 
\item
String theorists construct models of theories with complex and 
degenerate vacuum states in higher dimensions.
\end{enumerate}

\begin{figure}[tb]   
\begin{center}
\includegraphics[width=0.7\linewidth]{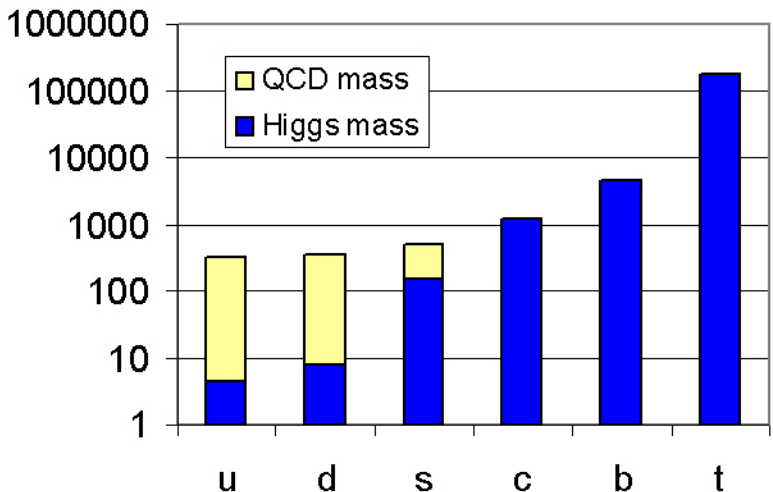}
\end{center}
\caption{Masses of the six known flavors of quarks (in MeV). The 
         mass generated by the Higgs vacuum is shown in blue, the
         mass generated by the QCD vacuum is shown in taupe.}
\label{fig5}
\end{figure}

\begin{figure}[tb]   
\begin{center}
\includegraphics[width=0.8\linewidth]{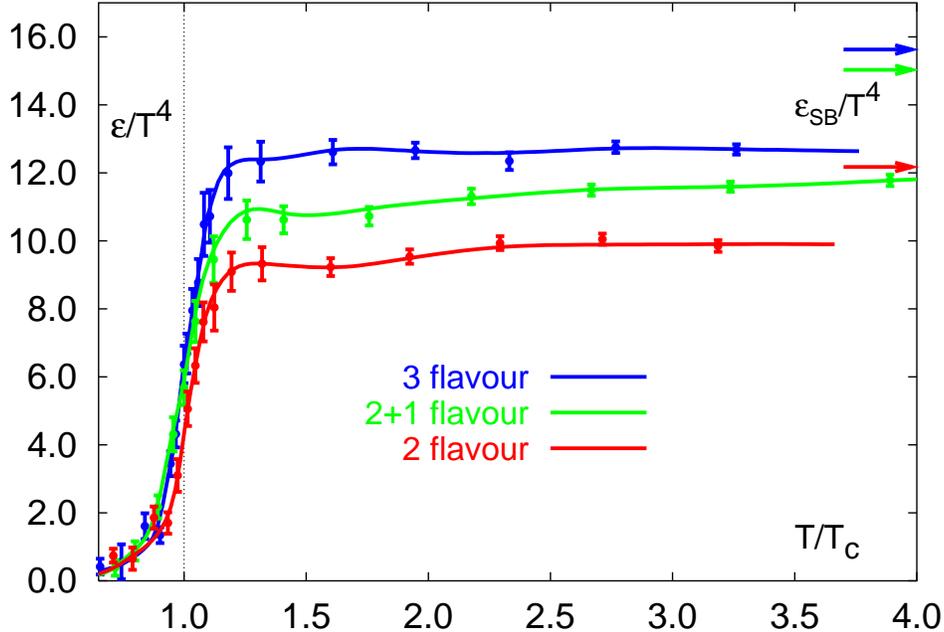}
\end{center}
\caption{Equation of state of hot strongly interacting matter.
         The ratio of the energy density $\epsilon$ to the fourth
         power of the temperature $T$ is a measure of the effective
         number of degrees of freedom. The rapid rise near $T=T_c$
         indicates the liberation of colour at the transition from
         a gas of hadrons to a quark-gluon plasma (from Karsch
         \cite{Ka02}).}
\label{fig6}
\end{figure}

The third approach, relativistic heavy ion collisions, alone 
allows to explore the notion that the properties of particles
and forces can be modified by a change in the vacuum state. 
The beam energies accessible at RHIC are high enough to create 
temperatures commensurate with the energy scale of the QCD
vacuum condensates and thus to affect the structure of the 
QCD vacuum. Numerical simulations of lattice QCD predict that
a dramatic change in the QCD vacuum state occurs around a
``critical'' temperature $T_c \approx 160$ MeV, where the 
quark and gluon condensates melt and the vacuum takes on a
trivial, perturbative structure. The part of the light quark 
masses that is induced by the quark condensate disappears 
above $T_c$ and only the much smaller mass generated by the
electroweak Higgs field remains (see Fig.~\ref{fig5}). The 
degrees of freedom corresponding to freely propagating gluons, 
which are frozen in the normal QCD vacuum, are also liberated 
above $T_c$. This pattern is illustrated in Fig.~\ref{fig6}, 
which shows the dramatic jump in the scaled energy density 
$\epsilon(T)/T^4$ at the critical temperature. 

Our universe began its life far above $T_c$ and cooled below
$T_c$ after about 20 $\mu$s. The physics program of the RHIC 
aspires to recreate this process in the collisions of two
$^{197}$Au nuclei with energies of up to 100 GeV per nucleon
each \cite{HM96}. In the laboratory, the two colliding nuclei 
appear highly Lorentz contracted (about 100-fold) as they
approach from opposite directions. As a consequence, the
immediate impact is extremely brief, leaving behind a highly
excited region of space, a sort of ``false vacuum'', as the 
debris from both nuclei recedes after the collision. The 
energy stored in the vacuum rapidly thermalizes due to the
nonlinear, chaotic dynamics of colour fields, producing a 
small region of highly heated vacuum roughly of the size
of a Au nucleus. Theoretical models, calibrated by the 
results from the first RHIC experiments, suggest that space 
is heated up to about 250 MeV, creating the proper conditions
for a change in the QCD vacuum. 

But how would we know that the predicted transformation 
actually occurs? Two effects caused by the disappearance of 
the vacuum condensates stand out: The melting of the quark 
condensate lowers the effective mass of the $s$-quark from 
about 500 MeV to less than 150 MeV, making it easy to create 
$s$-quark pairs in copious quantities \cite{RM82,KMR86}. The 
dissolution of the gluon condensate allows gluons to exist 
as abundant thermal excitations on which energetic partons 
can scatter and degrade their energy \cite{GW94,BSZ00}. 

These considerations suggest two critical signatures of the
predicted structural change in the QCD vacuum: Particles 
containing strange quarks should be produced in equilibrium
abundances, and particles created by fragmentation of high 
energy quarks should be substantially suppressed compared 
with naive expectations, an effect that has become known 
as ``jet quenching'' \cite{GW92}. Both predicted phenomena 
have, indeed, been observed in the RHIC experiments, which
we will review next.

\section{Results from RHIC}

\begin{figure}[tb]   
\begin{center}
\includegraphics[width=0.7\linewidth]{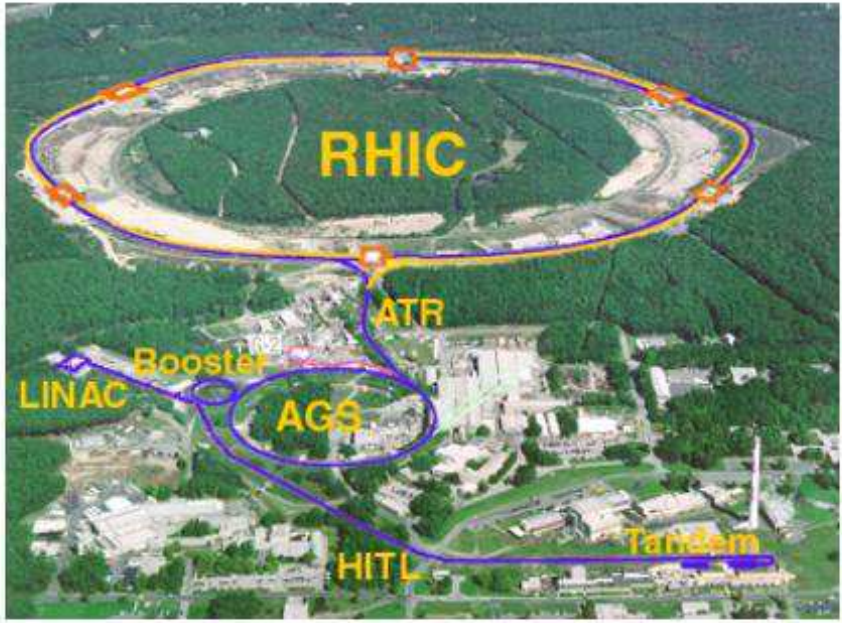}
\end{center}
\caption{The Relativistic Heavy Ion Collider (RHIC) accelerator
         complex at Brookhaven National Laboratory.}
\label{figR}
\end{figure}

The Relativistic Heavy Ion Collider, shown in Fig.~\ref{figR}, 
is an accelerator complex of great versatility. In addition 
to collisions between Au nuclei, it allows scientists to 
study collisions between two protons and even between heavy
hydrogen (deuterium) nuclei and Au nuclei. The latter serve
as benchmarks for the phenomena occurring in the normal QCD
vacuum at a given collision energy. The p+p and d+Au 
collisions are not expected to create a region of heated 
vacuum and thus should not exhibit the specific effects 
predicted to be signatures of a modified QCD vacuum state.
Since the year 2000, RHIC has had three major experimental runs
at collision energies of 130 GeV and 200 GeV per nucleon 
pair and involving all three systems mentioned above. A 
fourth run with colliding Au beams has just been completed.

\begin{figure}[tb]   
\begin{center}
\includegraphics[width=0.8\linewidth]{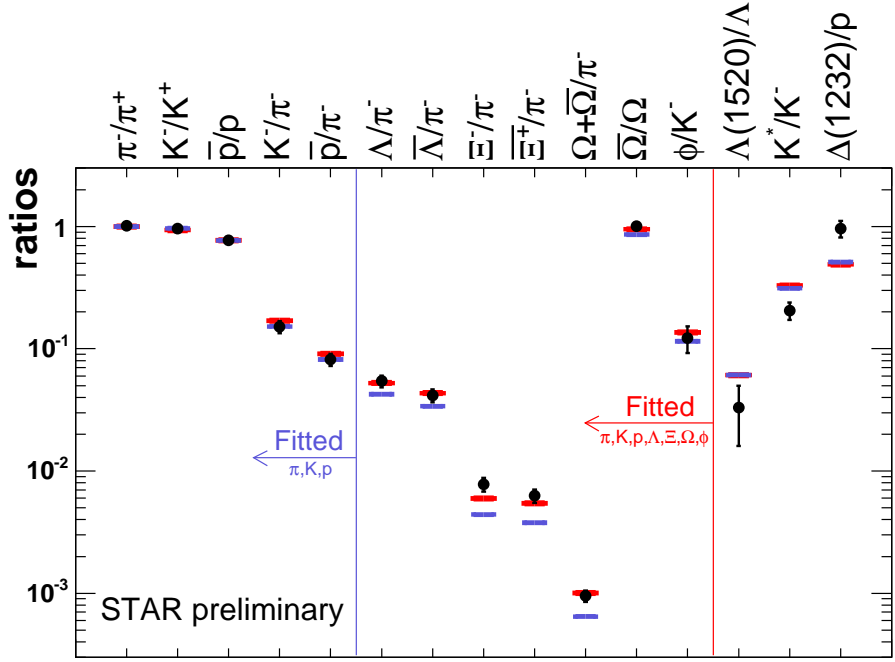}
\end{center}
\caption{Ratios of the yields of various hadrons emitted from
         a Au+Au collision at RHIC, in comparison with the
         predictions of a chemical equilibrium model.}
\label{fig7}
\end{figure}

\begin{figure}[tb]   
\begin{center}
\includegraphics[width=0.8\linewidth]{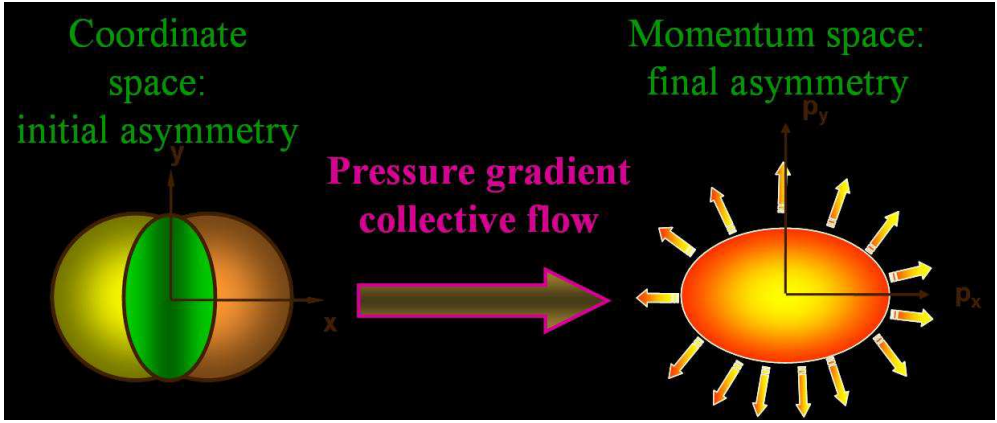}
\end{center}
\caption{The anisotropic shape of the hote fireball created in
         noncentral collisions of two nuclei results in an
         axially asymmetric pressure gradient which, in turn,
         leads to an elliptic anisotropy of the collective
         outward flow of the hot matter.}
\label{fig8}
\end{figure}

\begin{figure}[tb]   
\begin{center}
\includegraphics[width=0.45\linewidth]{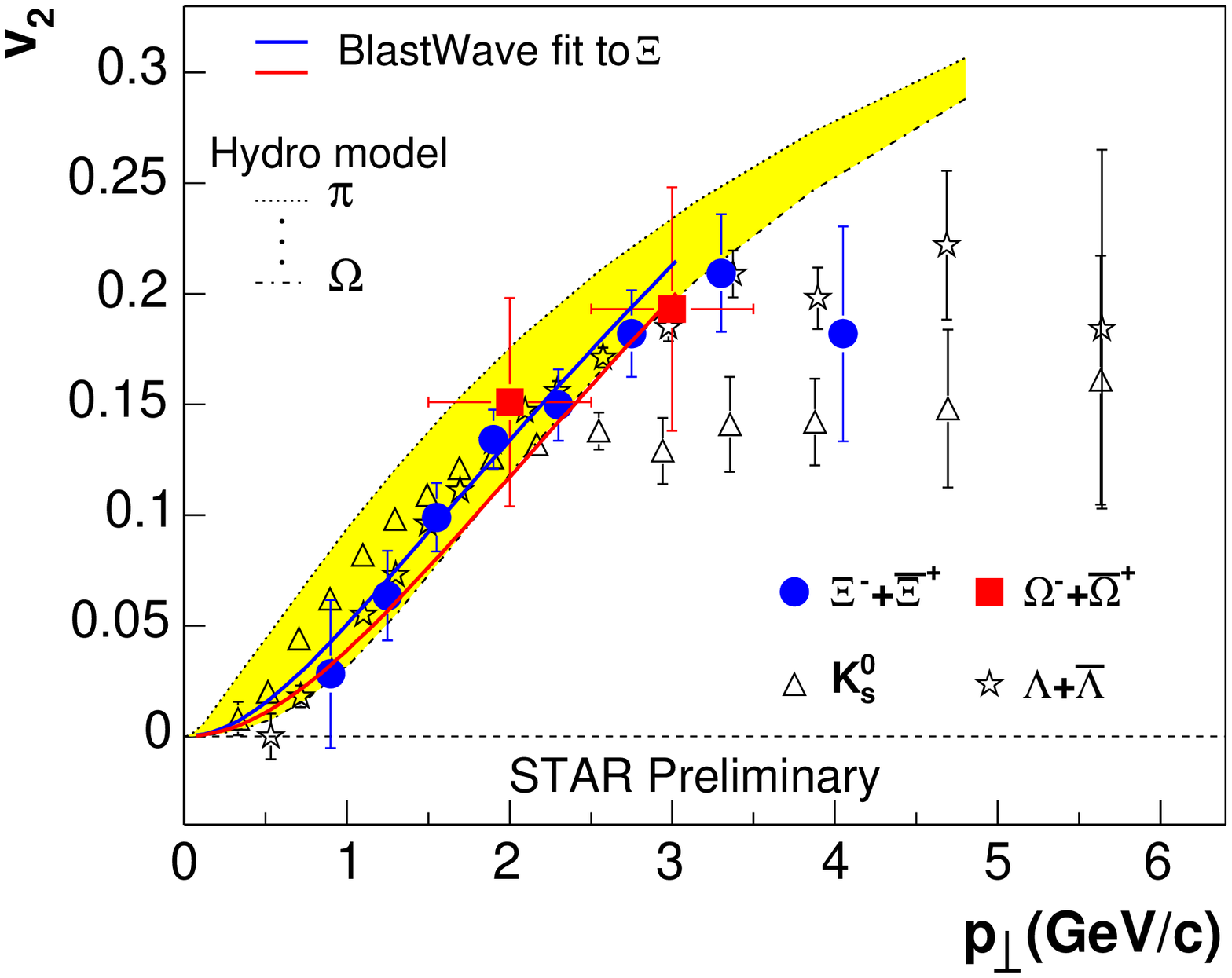}
\hspace{0.05\linewidth}
\includegraphics[width=0.45\linewidth]{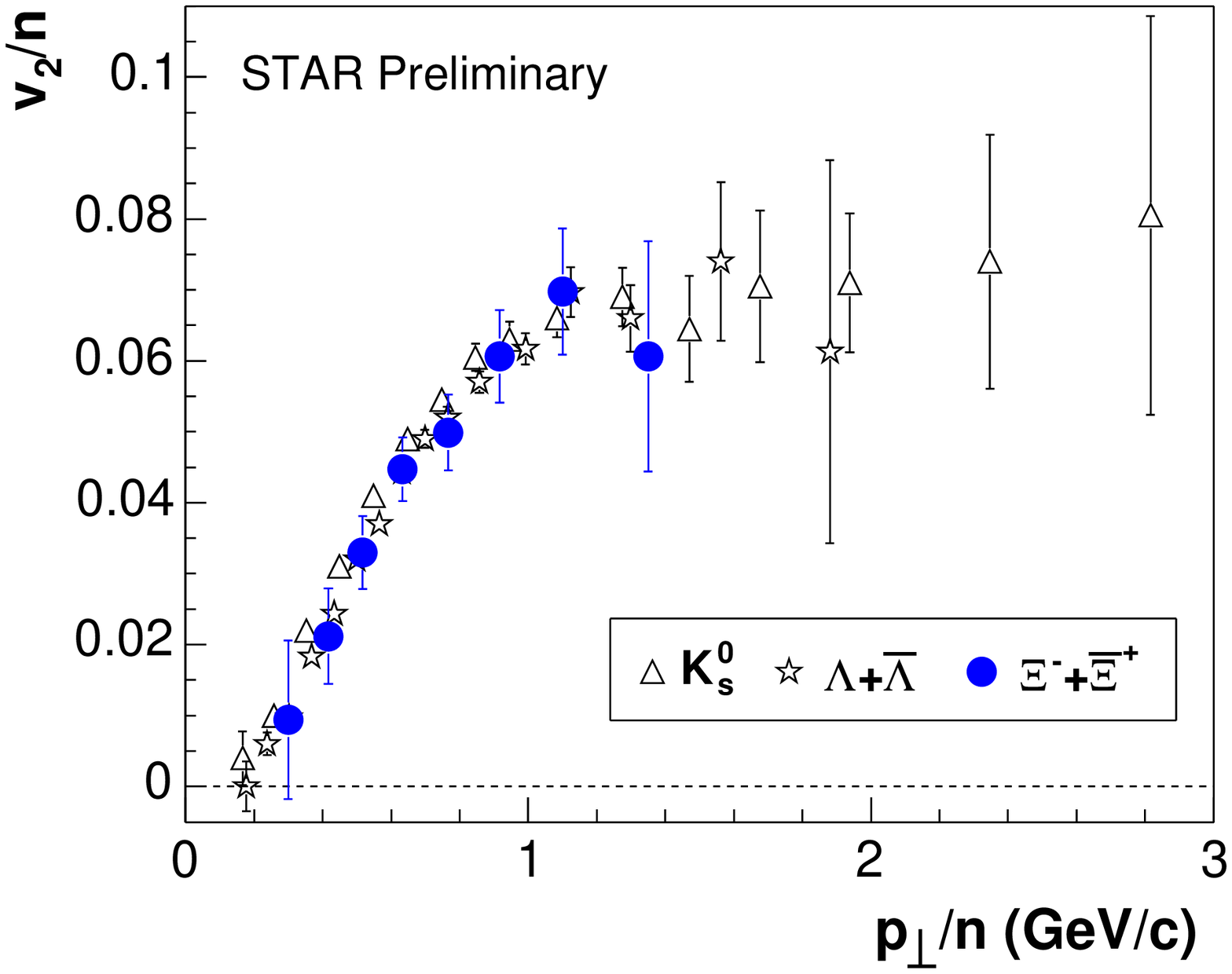}
\end{center}
\caption{The elliptic flow parameter $v_2$ of hadrons as function
         of the trannsverse momemtum differs by hadron species
         (left figure). The different dependences collapse into
         a single one, if the elliptic flow per constituent quark
         is plotted as a function of the quark transverse momentum
         (right figure).}
\label{fig9}
\end{figure}

Fig.~\ref{fig7} shows that hadrons containing any number
of $s$-quarks, up to three, are produced according to the
expectation that a quark-gluon plasma converts into hadrons 
near the critical temperature. The value deduced from the data 
($T = 160\pm 6$ MeV) is equal to $T_c$ within the theoretical 
uncertainties \cite{STAR-ch}. Measurements of the flow patterns 
of the emitted hadrons provide additional evidence that they 
are created directly from a phase of independent quarks and 
antiquarks. The experiment makes use of the fact that the
region of heated vacuum is almond-shaped in semiperipheral 
collisions between two nuclei, as illustrated in Fig.~\ref{fig8},
leading to an anisotropic expansion pattern, called ``elliptic 
flow'' \cite{Ol92}, and characterized by a parameter $v_2$. The 
observed flow anisotropy is a function of the momentum of the 
emitted particles and varies from one hadron species to another 
(see Fig.~\ref{fig9}, left). The remarkable finding is that all 
flow patterns collapse onto a common line when they are plotted 
per constituent quark in each hadron (see Fig.~\ref{fig9}, right), 
suggesting that one observes the flow pattern of individual 
quarks, which coalesce into hadrons only later \cite{v2}.

Basic principles of QCD mandate that the number of hadrons
emitted at large momentum transverse to the beam axis should 
grow like the number of collisions between pairs of nucleons, 
when two nuclei collide. As we argued above, the scattered 
partons, from which these hadrons are produced, are expected 
to suffer a substantial energy loss on the way out due to
collisions with thermal gluons, if the QCD vacuum state is 
altered. In QCD, the main mechanism for collisional energy 
loss is the radiation of a gluon by the struck parton, and
this effect is predicted to grow in proportion to the length
of the hot region traversed by the parton \cite{Ba97,BSZ00}. 
Because the yield of scattered partons falls rapidly as a 
function of the transverse momentum, this energy loss 
translates into a reduction in the yield at a fixed momentum. 
In fact, model calculations predict that most hadrons observed 
at a given momentum originate from partons scattered near the 
surfaces of the colliding nuclei \cite{Mu03}.

\begin{figure}[tb]   
\begin{center}
\includegraphics[width=0.6\linewidth]{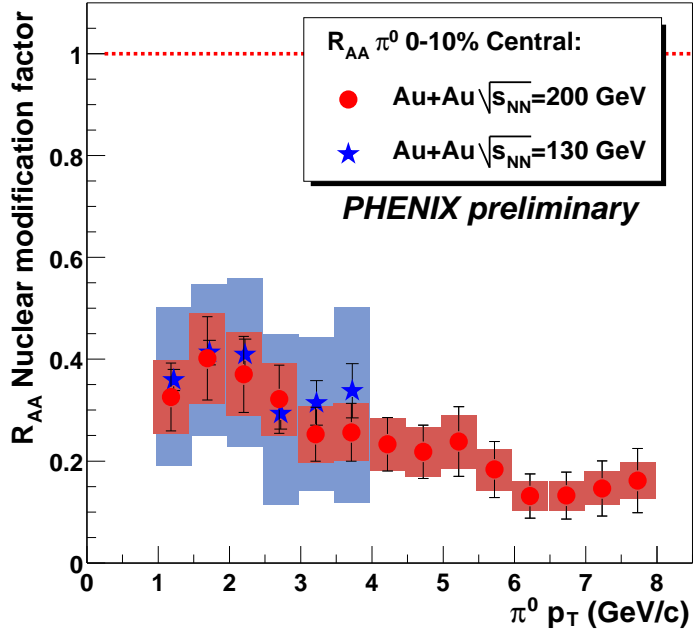}
\end{center}
\caption{Nuclear suppression of neutral pions observed in 
         $\sqrt{s_NN}=$200 GeV collisions of two Au nuclei
         (from PHENIX collaboration).}
\label{fig10}
\end{figure}

\begin{figure}[tb]   
\begin{center}
\includegraphics[width=0.45\linewidth]{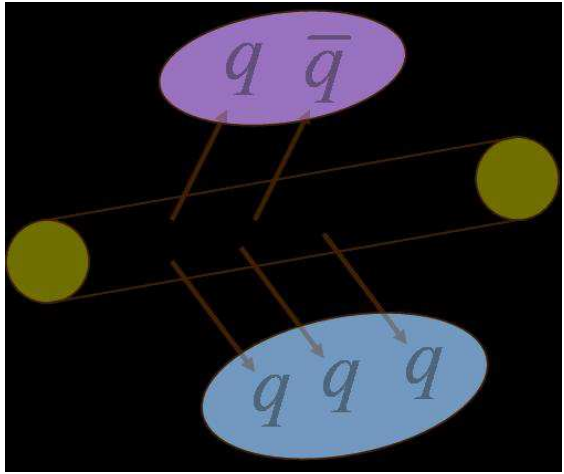}
\end{center}
\caption{Hadrons can be formed by recombination of quarks 
         from a dense system of deconfined partons. A quark
         and an antiquark form a meson, three quarks form a
         baryon.}
\label{fig11}
\end{figure}

The RHIC experiments clearly show the predicted suppression
of high energy hadrons (see Fig.~\ref{fig10}). Neutral pions 
with transverse momenta above 4 GeV/c are emitted five times 
more rarely in head-on collisions of two Au nuclei than 
extrapolated from the yield measured in $p+p$ collisions or 
grazing Au+Au collisions \cite{Jets}. This amount of suppression 
is consistent with that expected for a plasma containing free 
gluons at temperatures above $T_c$. The suppression effect 
is not observed in d+Au collisions, ruling out any 
initial-state effect associated with the properties of Au 
nuclei \cite{d+Au}. Again, a difference in the behaviour 
of mesons and baryons helps to better understand what is 
going on: The suppression of baryon production only sets in 
at higher momenta, above 5 GeV/c, because the coalescence 
of three quarks produces more energetic baryons than mesons,
which contain only a single quark pair, as illustrated in
Fig.~\ref{fig11} \cite{p-pi,v2,Fr03}.

To confirm the interpretation of these results, two other 
effects will be explored, which are sensitive to the density
of gluons in the hot QCD plasma. The first one makes use of
the fact that a dense plasma of gluons screens the normally 
long-ranged colour force. The strength of the screening effect 
is measured by the inverse screening length $\mu$, which is 
a function of the temperature $T$. Lattice simulations predict 
that the screening effect abruptly disappears as $T$ approaches 
$T_c$ from above \cite{Kac00}, indicating the
transition from a trivial QCD vacuum state to one filled with 
a gluon condensate. The idea is to detect the colour screening 
effect via the disappearance of bound states of a pair of heavy 
quarks, such as $c\bar c$ or $b\bar b$ \cite{MS86}. These states 
are denoted as $J/\Psi$ and $\Upsilon$ states, respectively. 
The recent high-statistics Au+Au run of RHIC is expected to allow 
for a precise measurement of the change in the $J/\Psi$ yield 
compared with extrapolations from $p+p$ collisions.

Another possible method of measuring the gluon density is 
by detecting energetic photons. These can either be created 
in the initial impact between the two nuclei, or during the
passage of scattered, energetic quarks through the hot plasma
\cite{FMS,BMS-phot}. The first process can be accurately 
predicted by extrapolation from $p+p$ collisions, because 
the photons are not affected by the change in the QCD vacuum. 
A precise measurement of the yield of energetic photons, 
which is difficult due to the presence of a large background 
of secondary photons from meson decays, would permit an 
independent determination of the gluon density.

\section{Summary and Outlook}

Through a series of observations and plausibity arguments, 
modern cosmology has led scientists to seriously consider 
the spectre of our universe as one among a vast multitude 
of universes, in which the laws of nature take on different 
forms. The inflationary cosmological model, which predicts 
an unlimited number of universes emerging in random 
succession from their individual Big Bang, is increasingly 
well supported by astrophysical evidence. The best way to 
reconcile the observed immutability of the laws of nature 
in our universe over the whole range of visible space and 
time with their posited variability from one cosmos to 
another is to consider this as an effect of the variability 
of the vacuum state. 

Experiments with relativistic heavy ions allow us, for the 
first time, to verify that a vacuum state can be modified, 
causing dramatic changes in the properties of fundamental 
particles (quarks and gluons) and the strong forces between 
them. The theoretical tools developed to understand this 
process in detail will also help us to explore models of vacuum 
variability affecting other particles and forces, as suggested 
by cosmology. 

\section*{\it Acknowledgements}

This work was supported in part by a research grant from the
U.~S.~Department of Energy. We thank T.~Ludlam for providing
an aerial view picture of the RHIC accelerator complex at
Brookhaven National Laboratory.


\begin{thebibliography}{99}
\bibitem{SUSY}
This estimate assumes that the physical vacuum energy is
determined by the scale at which supersymmetry is broken,
$\Lambda \approx M_{\rm SUSY}^4 \ge (1 {\rm TeV})^4$.
\end{thebibliography}
\end{document}